\documentclass[twocolumn,aps,pra,showpacs,amsmath,amssymb,superscriptaddress]{revtex4}

\usepackage{graphicx}
\usepackage{dcolumn}
\usepackage{bm}
\usepackage{hyperref}
\usepackage{rotating}
\def\<{\langle}
\def\>{\rangle}
\def\Ftb{F_{2{\rm b}}}
\def\Fthb{F_{3{\rm b}}}
\def\Mftb{M_{F_{2{\rm b}}}}
\def\Mfthb{M_{F_{3{\rm b}}}}

\graphicspath{{converted_graphics/}}
\begin{document}

\preprint{APS/123-QED}
\title{Universal few-body physics in resonantly interacting spinor condensates}

\author{V. E. Colussi}
\address{Department of Physics, University of Colorado, Boulder, Colorado 80309-0440, USA}
\author{Chris H. Greene}
\address{ Department of Physics, Purdue University, West Lafayette, Indiana 47907-2036, USA}
\author{J. P. D'Incao}
\address{Department of Physics, University of Colorado, Boulder, Colorado 80309-0440, USA}
\address{ JILA, University of Colorado and NIST, Boulder, Colorado 80309-0440, USA}

\begin{abstract}

Optical trapping techniques allow for the formation of bosonic condensates with internal degrees of freedom, so-called spinor condensates.  
Mean-field models of spinor condensates highlight the sensitivity of the quantum phases of the system to the relative strength of the two-body 
interaction in the different spin-channels.  Such a description captures the regime where these interactions are weak.  In the opposite and 
largely unexplored regime of strongly correlated spinor condensates, three-body interactions  can play an important role through the Efimov 
effect, producing possible novel phases. 
Here, we study the three-body spinor problem using the hyperspherical adiabatic representation for spin-1, -2 and -3 condensates 
in the strongly-correlated regime. We characterize the Efimov physics for such systems and the relevant three-body mean-field 
parameters. We find that the Efimov effect can strongly affect the spin dynamics 
and three-body mean-field contributions to the possible quantum phases of the condensate through universal contributions to scattering observables. 
\end{abstract}


\maketitle

\section{\label{sec:level1}Introduction\protect\\}
Over the last few decades technological advancements in trapping have allowed for the creation of a range of ultracold dilute gases in 
the quantum degenerate regime.  In bosonic systems, quantum statistics lead to a global phase coherence and macroscopic occupation
of the ground state of the system.  In spinor condensates the atomic spin degrees-of-freedom are energetically accessible leading to a 
unique scenario in which {\em both} superfluity and magnetism can exist \cite{SpinorReview}. The first spinor 
condensates were made with $^{87}$Rb in a double magnetic trap \cite{SpinorJILA} and with $^{23}$Na in an 
optical trap \cite{SpinorMIT}.  There has been a subsequent explosion of experimental 
\cite{SpinorExp1,SpinorExp2,SpinorExp3,SpinorExp4,SpinorExp5,SpinorExp6,SpinorExp7,SpinorExp8,SpinorExp9,SpinorExp10,SpinorExp11,SpinorExp12}, 
and theoretical \cite{SpinorTheo1,SpinorTheo1x,SpinorTheo2,SpinorTheo3,SpinorTheo4,SpinorTheo5,SpinorTheo6,SpinorTheo7,SpinorTheo8,SpinorTheo9,SpinorTheo10}
studies in the area. Studying the interplay between superfluidity and magnetism has proven to be a rich area for probing the many-body dynamics of the static and
nonequilibrium behavior of these systems leading to novel phenomena such as spin textures, spin domains, and complex spin mixing dynamics \cite{SpinorReview}. 

Many of the interesting phenomena in spinor condensates hinge on extreme sensitivity to the relative strengths of two-body interactions between the 
internal states; those interactions are typically weak for alkali atoms (with the exception of $^{85}$Rb \cite{klausen2001PRA,kempen2002PRL}, and possibly also $^{133}$Cs and $^7$Li) and as a consequence the spinor physics 
with strong interactions has largely remained unexplored. Although the usual Feshbach resonance technique for tuning the interactions \cite{chin2010RMP} cannot immediately be applied to spinor condensates, several proposals for achieving such control exist 
\cite{gerbier2006PRA,zhang2009PRL,hamley2009PRA,kaufman2009PRA,tscherbul2010PRA,hanna2010NJP,papoular2010PRA}, which enable possible
ways to study strongly correlated spinor condensates. Evidently, as the interactions become resonant, three-body contributions need
to be considered. 
In the strongly correlated regime, i.e., when the $s$-wave two-body scattering lengths associated with the hyperfine spin states 
exceed the typical range of interatomic interactions (the van der Waals length, $r_{\rm vdW}$), Efimov physics becomes important
\cite{efimov1970SJNP,braaten2006PR,wang2013AAMOP}, and several new aspects have to be considered. For instance, the existence
of an infinity of weakly bound Efimov trimers strongly affects the scattering observables at ultracold energies and can potentially impact the
spin dynamics in spinor condensates. Recent experimental and theoretical work on the three-body parameter 
\cite{IBK_Exps,LENS_Exps,Rice_Exps,Kayk_Exps,Jochim_Exps,Ohara_Exps,Ueda_Exps,JILA_Exps,Chin3BP,Ueda_Theo,JILA_Theo,Schimdt,Jensen}
has extended the universal properties of Efimov states allowing for further quantitative predictions of few-body phenomena relevant for spinor condensates.

One of the major differences between the usual ``spinless" Efimov physics and the effect for spinor systems
is the presence of multiple length scales in the problem. In spinor condensates the atomic energy levels are
$(2f+1)$-fold degenerate ($f$ is the atomic hyperfine angular momentum and $m_f=-f ...f$ its azimuthal component), and
there exist $f+1$ rotationally-invariant $s$-wave scattering lengths \cite{SpinorReview}. In Ref. \cite{Colussi2014} we explored in depth the Efimov physics for
$f=1$ spinor condensates and showed how the presence of the additional internal degrees of freedom can lead to
multiple families of Efimov states that strongly affect the condensate spin dynamics. In the context of nuclear physics, 
where isospin symmetry plays an important role, the work of Bulgac and Efimov \cite{bulgac1976SJNP} has also
demonstrated a much richer structure for Efimov physics when the spin degree of freedom was considered. As we will see later in 
this manuscript, a fundamental difference of the spin physics in nuclear systems and spinor condensates is that in a gas phase the atoms can be prepared in 
a single spin substate, which is not an eigenstate of well-defined total angular momentum.

Our treatment begins with a summary of the multichannel generalization of the adiabatic hyperspherical representation using contact two-body interactions.  
We then characterize the Efimov physics for spin-1, -2 and -3 by determining the topology of the three-body hyperspherical potentials and analysing
the possible families of Efimov states present in each case.  Special attention is given to the spin-2 $^{85}$Rb condensate due to the naturally large
values for the atomic scattering lengths found in Ref. \cite{klausen2001PRA,kempen2002PRL}. Finally, we determine the important three-body 
mean-field parameters,
thus building the fundamental blocks for future studies of three-body effects in spinor condensates.

\section{\label{sec:level2}Adiabatic hyperspherical representation for spinor systems}

Our study of few-body physics in spinor condensates begins from the multichannel generalization of the zero-range 
Fermi pseudopotential for s-wave interactions. In Refs.~ \cite{mehta2008PRA,rittenhouse2010PRA,rittenhouseThesis,KartavsevMacek},
a suitable generalization was proposed to be (in a.u.)
\begin{equation}
\hat{v}(r)=\frac{4\pi \hat{A}}{m}\delta^3(\vec{r})\frac{\partial}{\partial r}r,\label{eq:pseudo}
\end{equation}
where $\delta^3(\vec{r})$ is the three-dimensional Dirac delta-function, $m$ is the atomic mass, 
and $\hat{A}$ is a scattering length {\em operator} which incorporates all
the important multichannel structure of the two-body interactions \cite{mehta2008PRA}. 
For spinor condensates we assume the scattering length operator in Eq. (\ref{eq:pseudo}) can be defined for each pair of atoms as
\begin{equation}
\hat{A}=\sum_{\Ftb\Mftb}|\Ftb\Mftb\>a_{\Ftb}\<\Ftb\Mftb |,\label{eq:SpinoneAmatrix}
\end{equation}
where
\begin{align}
&|F_{\rm 2b}M_{F_{\rm 2b}} \rangle \equiv |(f_1f_2)F_{\rm 2b}M_{F_{\rm 2b}} \rangle=\nonumber\\
&\sum_{m_{f_{1}}m_{f_{2}}}\langle f_{1}m_{f_{1}}f_{2}m_{f_{2}}|F_{\rm 2b}M_{F_{\rm 2b}}\rangle|f_{1}m_{f_{1}}\rangle|f_{2}m_{f_{2}}\rangle.\label{2Spin}
\end{align} 
is the two-body spin functions of total angular momentum $|f_{1}-f_2|\le F_{\rm 2b} \le f_1+f_2$
and projection $M_{F_{\rm 2b}}=m_{f_1}+m_{f_2}$, expressed in terms
of the Clebsch-Gordan coefficients. Therefore, the two-body interaction in Eq.~(\ref{eq:pseudo}) is diagonal in the
spin basis $\{|\sigma\>\}=\{|F_{2{\rm b}}M_{F_{2{\rm b}}}\>\}$. We note that, due to bosonic symmetry only the symmetric 
spin states ($F_{\rm 2b}$ even) are allowed to interact with rotationally-invariant scattering lengths 
$a_{F_{\rm 2b}}\equiv\{a_0, a_2, ..., a_{2f}\}$.  
These scattering lengths set important length scales in the system.  Many-body properties of the system such as the miscibility of spin components are sensitive to the relative strength and sign of the scattering lengths \cite{SpinorReview}.  Their strength also signifies the appearance of universal three-body physics in the scattering observables.  

The three-body problem is solved using a multichannel generalization of the adiabatic hyperspherical method via a Green's function approach developed in Ref.~\cite{mehta2008PRA,rittenhouse2010PRA,rittenhouseThesis}.  As usual, the three-body system is characterized by a single length scale, the hyperradius 
$R=(\rho_1^2+\rho_{2}^2)^{1/2}$. The mass-scaled Jacobi vectors (written in the ``odd-man-out'' notation) are given in terms of the atomic distances by
\begin{eqnarray}
\vec{\rho}_{1}^{(k)}=(\vec{r}_{j}-\vec{r}_{i})/d, \mbox{~~~and~~~}
\vec{\rho}_{2}^{(k)}=d\left(\vec{r}_{k}
-\frac{\vec{r}_{i}+\vec{r}_{j}}{2}\right), \label{JacobiMass}
\end{eqnarray}
where $d=2^{1/2}/3^{1/4}$. A set $\Omega$ of six hyperrangles describe all other internal degrees of freedom associated to the system's internal motion. 
Treating the hyperradius as an adiabatic parameter, the three-body wave-function can be expressed as
\begin{eqnarray}
\Psi(R,\Omega)=\sum_{\nu}F_{\nu}(R)\sum_{\Sigma}\Phi_{\nu}^{\Sigma}(R;\Omega)|\Sigma\>
\end{eqnarray}
where $F(R)$ are the hyperradial wave functions, $\Phi(R;\Omega)$ the channel functions and 
$\{|\Sigma\>\}=\{|m_{f_1},m_{f_2},m_{f_3}\>\}$ the three-body spin functions, formed by the combination of all possible product states, chosen to 
simplify the formulation.  The adiabatic channel functions are eigenstates of the fixed-$R$ hyperangular Schr\"odinger equation 
\begin{align}
\sum_{\Sigma'}\Big[&\frac{\hat{\Lambda}^2(\Omega)+15/4}{2\mu R^2}\delta_{\Sigma\Sigma'}+\<\Sigma'|\hat{V}(R,\Omega)|\Sigma\>
\nonumber\\
&+{E_{\Sigma}}\delta_{\Sigma\Sigma'}\Big]\Phi_{\nu}^{\Sigma}(R;\Omega)=U_{\nu}(R)\Phi_{\nu}^{\Sigma}(R;\Omega),\label{eq:schrodinger}
\end{align}
with corresponding eigenvalues given by the adiabatic potentials $U(R)$ which determine the hyperradial motion and describe (in conjunction with the nonadiabatic couplings) the possible bound states and scattering properties of the system. In the above equation, $\mu=m/\sqrt{3}$ is the the three-body reduced mass, 
$\hat{\Lambda}$ is the grand angular momentum \cite{avery}, $\hat{V}$ is the sum of all possible pairwise interactions
[Eq.~(\ref{eq:pseudo})],
given by
\begin{eqnarray}
\hat{V}=\frac{4\pi}{m}\sum_{i<j}|f_km_{f_k}\rangle\left[\hat{A}^{(k)}\delta^3(d\vec\rho_1^{(k)})\frac{\partial }{\partial \rho_1^{(k)}}\rho_1^{(k)}\right]\langle f_km_{f_k}|,
\end{eqnarray}
where the superscript $(k)$ for the ``odd-man-out'' signifies that the interaction is between particles $i$ and $j$ while particle $k$ spectates.
In the present study the three-body energy levels, $E_{\Sigma}$, are degenerate and we set them to zero.  

Equation~(\ref{eq:schrodinger}) is solved using a hyperangular Green's function and the corresponding Lippmann-Schwinger 
equation \cite{mehta2008PRA} for each component of the channel function,  
\begin{align}
\Phi_\Sigma(R;\Omega)=-2\mu R^2\sum_{\Sigma',k}\int {\rm d}\Omega ' G_{\Sigma\Sigma}(\Omega,\Omega')\nonumber\\
\times v^{(k)}_{\Sigma\Sigma'}(R,\Omega')\Phi_{\Sigma'}(R;\Omega'),\label{eq:LS}
\end{align}
where 
\begin{equation}
v^{(k)}_{\Sigma\Sigma'}(R,\Omega)=\<\Sigma| v(r_{ij})|\Sigma'\>.\label{eq:pairwise}
\end{equation}
The components of the hyperangular Green's function satisfy $[\hat{\Lambda}^2-(s^2-4)]G_{\Sigma\Sigma}(\Omega,\Omega')=\delta(\Omega,\Omega')$, 
where $s$ is related to the hyperangular eigenvalue $U(R)$ through  
\begin{equation}
U(R)=\frac{s(R)^2-1/4}{2\mu R^2}.\label{eq:ThreeBodyPotential}
\end{equation}
Evaluating the integral over $v_{\Sigma\Sigma'}^{(k)}$, considering only states of total orbital angular momentum $L=0$, and solving the 
Lippmann-Schwinger equation reduces to determining values of $s$ for which the determinant of the matrix  
\begin{equation}
\hat{Q}=\left [\frac{3^{1/4}}{2^{1/2}R}\left(M^{(1)}+M^{(2)}P_-+M^{(3)}P_+\right)-1\right]\label{eq:Q}
\end{equation}
vanishes \cite{KartavsevMacek,mehta2008PRA}.  In the equation above the matrix element for the $M$ matrices are given by
\begin{eqnarray}
M^{(i)}_{\Sigma\Sigma'}=
\begin{cases} 
A_{\Sigma\Sigma'}^{(i)}s\cot(s\pi/2), & i=1, \\
-A_{\Sigma\Sigma'}^{(i)}\frac{4\sin(s\pi/6)}{\sqrt{3}\sin(s\pi/2)}, & i=2,3,
\end{cases}
\end{eqnarray}
where $A_{\Sigma\Sigma'}^{(i)}$ is the matrix element for the two-body scattering matrix written in the three-body spin basis,
\begin{align}
A^{(i)}_{\Sigma\Sigma'}&=\<\Sigma|\hat{A}^{(i)}|\Sigma'\>\nonumber\\
&=\sum_{\Ftb,\Mftb}  a_{F_{\rm 2b}}\<m_{f_j}m_{f_k}|\Ftb\Mftb\>\nonumber\\
&~~~~~~~~\times\<\Ftb\Mftb|m_{f_j}'m_{f_k}'\>\delta_{m_{f_i}m_{f_i}'}    \label{eq:Athreebodymatrix}
\end{align}
with the $i$th particle as spectator to the interaction.  
The $P_+$ and $P_-$ matrices represent cyclic and anti-cyclic permutations of the the three-body spin basis
\begin{eqnarray}
(P_+)_{\Sigma\Sigma'}&=&\<\Sigma |P_{123}|\Sigma'\>\nonumber\\
&=&\<m_{f_1},m_{f_2},m_{f_3}|P_{123}|m_{f_1}',m_{f_2}',m_{f_3}'\>\nonumber\\
&=&\delta_{m_{f_1},m_{f_2}'}\delta_{m_{f_2},m_{f_3}'}\delta_{m_{f_3},m_{f_1}'}\label{eq:Pplus}\\
(P_-)_{\Sigma\Sigma'}&=&\<\Sigma |P_{132}|\Sigma'\>\nonumber\\
&=&\<m_{f_1},m_{f_2},m_{f_3}|P_{132}|m_{f_1}',m_{f_2}',m_{f_3}'\>\nonumber\\
&=&\delta_{m_{f_1},m_{f_3}'}\delta_{m_{f_2},m_{f_1}'}\delta_{m_{f_3},m_{f_2}'}.\label{eq:Pminus}
\end{eqnarray}

The spin basis $\{|\Sigma\>\}$ was chosen to to give simple expressions for the components of $\hat{Q}$ which were 
written in the odd-man-out notation. In this form, however, $\hat{Q}$ does not account for any symmetry property in the system.  
This is fixed by the unitary transformation $\hat{S}\hat{Q}\hat{S}^T$ with 
\begin{align}
(S)_{\Sigma\Sigma'}=\<\Sigma|\left(\sum_{\Ftb} |\Fthb\Mfthb(\Ftb)\>\<\Fthb\Mfthb(\Ftb)|\right)|\Sigma'\>,\label{eq:Smatrix}
\end{align}
where the three-body spin functions of total angular momentum  $|F_{2{\rm b}}-f|\leq F_{3{\rm b}} \leq|F_{2{\rm b}}+f|$ and 
projection $M_{F_{3{\rm b}}}=M_{F_{2{\rm b}}}+m_f$ are given by 
\begin{align}
&|F_{\rm 3b}M_{F_{\rm 3b}}(F_{\rm 2b})\rangle\equiv |(f_1f_2f_3)F_{\rm 3b}M_{F_{\rm 3b}}(F_{\rm 2b})\rangle=\nonumber\\
&\sum_{M_{F_{\rm 2b}}m_{f_3}}
\langle F_{\rm 2b}M_{F_{\rm 2b}}f_3 m_{f_3}|F_{\rm 3b}M_{F_{\rm 3b}}\rangle
|F_{\rm 2b}M_{F_{\rm 2b}}\rangle|f_3m_{f_3}\rangle.\label{eq:threebodyspinfunctions}
\end{align}
Therefore, evaluating the det[$\hat{S}\hat{Q}\hat{S}^T$]=0 results in a transcendental equation whose roots 
$s(R)$ determine the channel functions $\Phi(R;\Omega)$ and the three-body potentials $U(R)$ from Eq.~(\ref{eq:ThreeBodyPotential})
with well define hyperfine angular momentum ($F_{\rm 3b}$ and $M_{F_{\rm 3b}}$) and permutation symmetry. 
In the usual spinless problem, the solution for three identical bosons in the limit $R/a\rightarrow0$ ($a$ been the scattering length)
gives a lone imaginary root $s_0\approx1.0062i$, which produces an attractive three-body potential when substituted in Eq.~(\ref{eq:ThreeBodyPotential}). 
This potential supports an infinite number of bound trimers characteristic of the Efimov effect. In the spinor case elaborated above, the degeneracy 
of the hyperfine spin-manifold yields fundamentally different three-body physics as demonstrated in the following two sections.

\section{\label{sec:level3}Spin-$1$ systems}

\begin{figure}[htbp]
\includegraphics[width=3.4in]{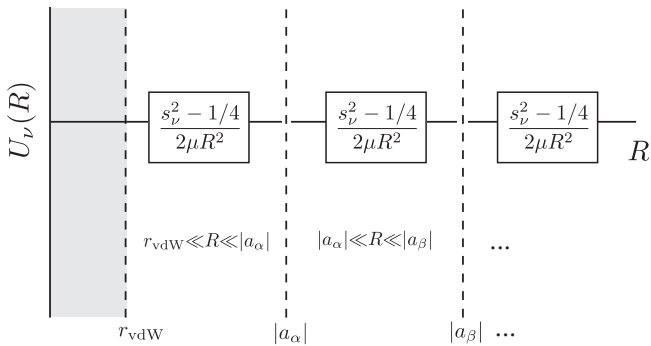}
\caption{Schematic representation of the relevant regions in $R$ of the three-body potentials $U(R)$. The shaded area, $R<r_{\rm vdW}$,
represents the region where the three-body potentials are expected to be non-universal. For all other regions, the potentials are universally
given by the form in Eq.~(\ref{eq:ThreeBodyPotential}) with $s(R)\equiv s$. \label{SchematicPotentials}}
\end{figure}

In this section, the results for spin-1 from Ref.~\cite{Colussi2014} are summarized, focusing on the structure of the three-body potentials and 
the families of Efimov states. The impact of Efimov physics on the three-body scattering observables and the many-body physics is a subject 
which is delayed until Sec.~V. In order to explore the solutions $s$ of det[$\hat{S}\hat{Q}\hat{S}^T$]=0 and the physics resulting from them, we 
assume the relevant scattering lengths for the problem, $a_{F_{\rm 2b}}\equiv\{a_0, a_2, ..., a_{2f}\}$, to differ greatly in magnitude from each other. 
By doing so, the values for $s$ with a range, say, $|a_\alpha|\ll R\ll |a_{\beta}|$ is constant and the corresponding values can be easily listed and
emphasize the attractive ($s$ imaginary) or repulsive ($s$ real) $1/R^2$ interaction that characterizes the Efimov physics.  When this condition of highly different scattering length magnitudes is not satisfied, one must in general return to the exact solutions of the above transcendental equation.
In the regimes considered here, there will exist typically various regions in which a detailed analysis must be performed (see Fig.~\ref{SchematicPotentials}).
Here this analysis is carried out for every possible case.

Following our analysis in the previous section we can derive the transcendental equations that determine $s_\nu$,
and consequently the three-body potentials $U_\nu(R)$ [see Eq.~(\ref{eq:ThreeBodyPotential})]. For $f=1$ atoms, 
the possible values for $F_{\rm 3b}$ are 0, 1, 2 and 3. An analysis of the three-body spin functions (\ref{eq:threebodyspinfunctions}) 
for $F_{\rm 3b}=0$ \cite{Colussi2014}, however, shows that such states are fully anti-symmetric and are non-interacting. [The
potential model in Eq. (\ref{eq:pseudo}) only captures $s$-wave interactions.] For $F_{\rm 3b}=1$, however, the spin functions 
are fully symmetric and the values for $s$ are determined by solving the transcendental equation,
\begin{align}
&\frac{3^{\frac{1}{4}}(a_0+a_2)s\cot(\frac{\pi}{2}s)}{2^{\frac{1}{2}}R}
-\frac{3^{\frac{1}{2}}(a_0a_2)s^2\cot(\frac{\pi}{2}s)^2}{2R^2}\nonumber\\
&-\frac{2^{\frac{3}{2}}(2a_0+a_2)\sin(\frac{\pi}{6}s)}{3^{\frac{5}{4}}\sin(\frac{\pi}{2}s)R}
+\frac{2~a_0a_2s\cot(\frac{\pi}{2}s)\sin(\frac{\pi}{6}s)}{\sin(\frac{\pi}{2}s)R^2}\nonumber\\
&+\frac{16~a_0a_2\sin(\frac{\pi}{6}s)^2}{3^{\frac{1}{2}}\sin(\frac{\pi}{2}s)^2R^2}=1. \label{EqF3b=1}
\end{align}
(Imaginary roots can be obtained by mapping $s\rightarrow is$.)
The above transcendental equation depends on both two-body scattering lengths, $a_{0}$ and $a_2$. 
For $F_{\rm 3b}=2$ and 3 the spin states are symmetric and mixed-symmetry states \cite{Colussi2014}, respectively, and the interaction strength $s$ is obtained, 
respectively, through,
\begin{eqnarray}
\frac{3^{1/4}a_2s\cot(\frac{\pi}{2}s)}{2^{1/2}R}
+\frac{2~2^{1/2}a_2s\sin(\frac{\pi}{6}s)}{3^{1/4}\sin(\frac{\pi}{2}s)R}=1,\label{EqF3b=2}
\end{eqnarray}
and
\begin{eqnarray}
\frac{3^{1/4}a_2s\cot(\frac{\pi}{2}s)}{2^{1/2}R}
+\frac{4~2^{1/2}a_2s\sin(\frac{\pi}{6}s)}{3^{1/4}\sin(\frac{\pi}{2}s)R}=1. \label{EqF3b=3}
 \end{eqnarray}
Note that for $F_{\rm 3b}=2$ and 3 the transcendental equations depend only on $a_{2}$. 
Table \ref{TabRoots2} lists the solutions of Eqs.~(\ref{EqF3b=1})--(\ref{EqF3b=3}) for the regions
in $R$ in which the value of $s$ is constant. Note that we also list the relevant scattering lengths for each value of $F_{\rm 3b}$.
For the cases where an imaginary root exists, i.e., when Efimov states are allowed, we list both the 
imaginary root and the lowest real root. For cases in which an imaginary root does not exist, we list only the lowest real root.
It is interesting noting that when $R$ is much smaller than all relevant scattering lengths, $R\ll |a_{\{0,2\}}|$, the imaginary root,
$s_0\approx1.0062i$, is the same than the usual three identical boson problem. For the regions $|a_0|\ll R\ll |a_2|$ and 
$|a_2|\ll R\ll |a_0|$, i.e., when the $a_0$ and $a_2$ are not effectively resonant, respectively, the possible roots can be
different than the ones obtained for the usual bosonic problem. These new roots result from the fact that the spin functions 
[Eq.~(\ref{eq:threebodyspinfunctions})] are superpositions of single particle product states (as determined by the 
angular momentum addition algebra). It is also interesting noting that for all cases we have explored, mixed-symmetry states 
($F_{\rm 3b}=2$ for $f=1$) display a stronger repulsive barrier whenever $R$ is larger than all scattering lengths ($s=4$) in 
comparison to symmetric states ($s=2$). This
implies that some three-body scattering observables should be suppressed for mixed-symmetry states. This will, in fact, simplify
some of our analysis in Section V.
\begin{figure*}[htbp]
\includegraphics[width=7in]{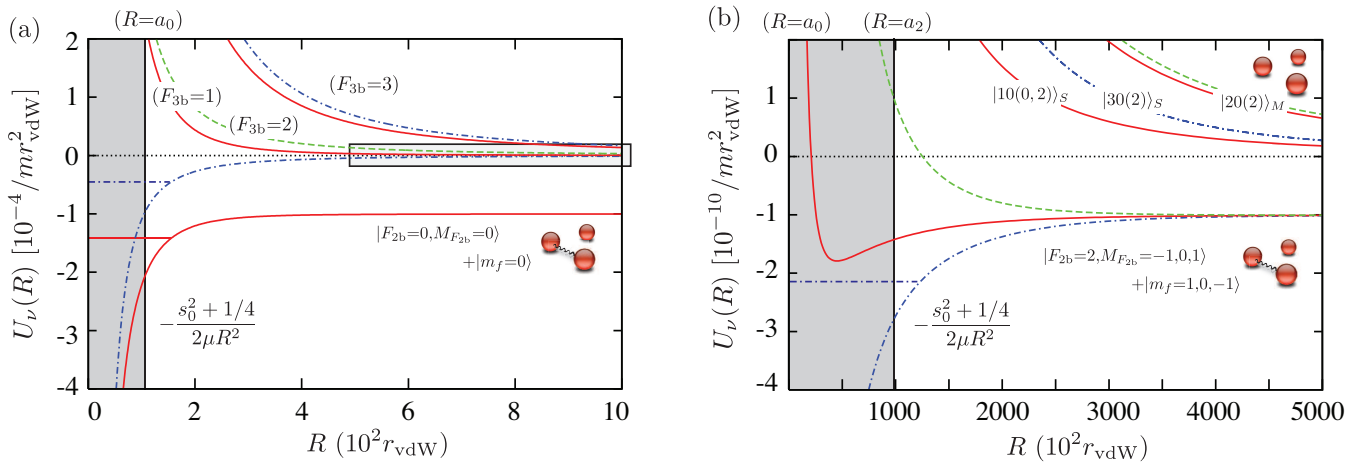}
\caption{$\Fthb=1$ (red solid line), 2 (green dashed line), and 3 (blue dash-dotted line) hyperspherical adiabatic potentials for $f=1$ atoms 
with $a_0=10^2r_{\rm vdW}$ and $a_2=10^5r_{\rm vdW}$.  (a) For $R\leq\{a_0,a_2\}$ (shaded region) two attractive potentials exist 
(both with $s_0\approx 1.0062i$), allowing for two families of Efimov states, and for $R>a_0$, 
one of these potentials turns into an atom-dimer channel $|\Ftb=0,\Mftb=0\> |f=1,m_f=0\>$.  (b) For $a_0\leq R\leq a_2$ 
(shaded region), only one family of Efimov states exists ($s_0\approx 1.0062i$), and for $R\gg a_2$ three 
(asymptotically degenerate) potentials describe atom-dimer channels, $|\Ftb=2,\Mftb=-1,0,1\>|f=1,m_f=1,0,-1\>$.\label{Potentials_f=1}}  
\end{figure*}

\begin{table*}[htbp]
\caption{Values of $s_{\nu}$ relevant for $f$$=$$1,2$ and $3$ spinor condensates covering all possible regions of $R$  and for different magnitudes of the relevant scattering lengths. We list the lowest few values of $s_{\nu}$ for each $F_{\rm 3b}$ and their multiplicity (superscript) if greater than one.
}\label{TabRoots2}
\begin{ruledtabular}
{\scriptsize 
\begin{tabular}{lccccccc}
 {($f=1$)}  & $F_{\rm 3b}=1$ &$ F_{\rm 3b}=2$ & $F_{\rm 3b}=3$ \\
 Relevant $a_{\Ftb}$& $\{a_0,a_2\}$&$\{a_{2}\}$&$\{a_{2}\}$\\
[0.05in]\hline
$R$$\ll$$|a_{\{0,2\}}|$  & $1.0062i,2.1662$ & $2.1662$ & $1.0062i,4.4653$ \\
$|a_0|$$\ll$$R$$\ll$$|a_2|$   & $0.7429$  & $2.1662$ & $1.0062i,4.4653$ \\
 $|a_2|$$\ll$$R$$\ll$$|a_0|$  & $0.4097$  &  $4$ & $2$ \\
$R$$\gg$$|a_{\{0,2\}}|$  & $2$  &  $4$  & $2$ \\
 [0.05in]\hline\\
{($f=2$)}  &$F_{\rm 3b}=0$ &$F_{\rm 3b}=1$ &$F_{\rm 3b}=2$ &$ F_{\rm 3b}=3$ &$F_{\rm 3b}=4$ &$F_{\rm 3b}=5$ &$F_{\rm 3b}=6$ \\
Relevant $a_{\Ftb}$&$\{a_{2}\}$&$\{a_2\}$&$\{a_{0},a_2,a_4\}$&$\{a_2,a_4\}$&$\{a_2,a_4\}$&$\{a_4\}$&$\{a_4\}$\\
[0.05in]\hline
$R$$\ll$$|a_{\{0,2,4\}}|$                   & $1.0062i, 4.4653$ &$ 2.1662 $& $1.0062i,2.1662^{(2)}$ & $1.0062i, 2.1662$          & $1.0062i, 2.1662$ & $2.1662$ & $1.0062i, 4.4653$ \\
$|a_0|$$\ll$$R$$\ll$$|a_{\{2,4\}}|$ & $1.0062i, 4.4653 $& $2.1662$ & $0.49050$                       & $1.0062i,2.1662 $          & $1.0062i, 2.1662$  & $2.1662$ & $1.0062i, 4.4653$\\
$|a_2|$$\ll$$R$$\ll$$|a_{\{0,4\}}|$ & $2$                           & $4$           & $0.7473i, 2.1662$          & $1.1044$                          & $0.66080$                     & $2.1662$ & $1.0062i, 4.4653$\\
$|a_4|$$\ll$$R$$\ll$$|a_{\{0,2\}}|$ & $1.0062i, 4.4653$ & $2.1662$ & $0.3788i, 2.1662$          & $0.5528i,3.5151$           & $0.52186$                    & $4$ & $2$\\
$|a_{\{0,2\}}|$$\ll$$R$$\ll$$|a_4|$ & $2$                           & $4$           & $0.97895$                       & $1.1044$                          & $0.66080$                     & $2.1662$ & $1.0062i, 4.4653$\\
$|a_{\{0,4\}}|$$\ll$$R$$\ll$$|a_2|$ & $1.0062i, 4.4653$ & $2.1662$ & $1.3173$                         & $0.5528i, 3.5151$           & $0.52186$                    & $4$ & $2$\\
$|a_{\{2,4\}}|$$\ll$$R$$\ll$$|a_0|$ & $2$                           & $4$           & $0.68609$                       & $2$                                     & $2$                   & $4$ & $2$\\
$R$$\gg$$|a_{\{0,2,4\}}|$                & $2$                           & $4$          &          $2$                                  & $2$                                           &$2$                        &$4$   &$2$\\
[0.05in]\hline\\
 {($f=3$)}  &$F_{\rm 3b}=1$ &$F_{\rm 3b}=2$ &$F_{\rm 3b}=3$ &$ F_{\rm 3b}=4$ &$F_{\rm 3b}=5$\\
 Relevant $a_{\Ftb}$& $\{a_2,a_4\}$&$\{a_2,a_4\}$&$\{a_0,a_2,a_4,a_6\}$&$\{a_2,a_4,a_6\}$&$\{a_2,a_4,a_6\}$\\
[0.05in]\hline
$R$$\ll$$|a_{\{0,2,4,6\}}|$   &$1.0062i,2.1662$&$2.1662^{(2)}$&$1.0062i^{(2)},2.1662^{(2)}$&$1.0062i,2.1662^{(2)}$&$1.0062i,2.1662^{(2)}$   \\
$|a_0|$$\ll$$R$$\ll$$|a_{\{2,4,6\}}|$  &$1.0062i,2.1662$&$2.1662^{(2)}$&$1.0062i,0.3420$&$1.0062i,2.1662^{(2)}$&$1.0062i,2.1662^{(2)}$ \\
$|a_2|$$\ll$$R$$\ll$$|a_{\{0,4,6\}}|$  &$0.6608$&$2.1662$&$1.0062i,0.5469$&$0.8754$&$0.2588$                       \\
$|a_4|$$\ll$$R$$\ll$$|a_{\{0,2,6\}}|$  &$0.5219$&$2.1662$&$1.0062i,1.1901$&$0.9352i,2.1662$&$0.6678i,2.1662$\\
$|a_6|$$\ll$$R$$\ll$$|a_{\{0,2,4\}}|$  &$1.0062i,2.1662$&$2.1662^{(2)}$&$1.0062i,0.4112i$&$0.4309i,2.1662$&$0.3351i$\\ 
$|a_{\{0,2\}}|$$\ll$$R$$\ll$$|a_{\{4,6\}}|$    &$0.6608$&$2.1662$&$0.6521i,1.0098$&$0.8754$&$0.2588$               \\
$|a_{\{0,4\}}|$$\ll$$R$$\ll$$|a_{\{2,6\}}|$    &$0.5219$&$2.1662$&$0.6080i,1.6045$&$0.9352i,2.1662$&$0.6678i,2.1662$\\
$|a_{\{0,6\}}|$$\ll$$R$$\ll$$|a_{\{2,4\}}|$       &$1.0062i,2.1662$&$2.1662$&$0.9666i,1.1343$&$0.4309i,2.1662$&$0.3351i,2.1662$ \\
$|a_{\{2,4\}}|$$\ll$$R$$\ll$$|a_{\{0,6\}}|$       &$2$&$4^{(2)}$&$0.5858i,1.7691$&$1.0111$&$0.9552$                           \\
$|a_{\{2,6\}}|$$\ll$$R$$\ll$$|a_{\{0,4\}}|$        &$0.6608$&$2.1662$&$0.9100i,1.1558$&$1.6693$&$1.2189$                       \\
$|a_{\{4,6\}}|$$\ll$$R$$\ll$$|a_{\{0,2\}}|$       &$0.5219$&$2.1662$&$0.4289i,1.2015$&$0.0803i,3.3774$&$0.8199$                            \\
$|a_{\{0,2,4\}}|$$\ll$$R$$\ll$$|a_6|$               &$2$&$4$&$0.9984$&$1.0111$&$0.9552$          \\
$|a_{\{0,2,6\}}|$$\ll$$R$$\ll$$|a_4|$                 &$0.6608$&$2.1662$&$0.6415i,2$&$1.6693$&$1.2189$          \\
$|a_{\{0,4,6\}}|$$\ll$$R$$\ll$$|a_2|$                &$0.5219$&$2.1662$&$0.6392$&$0.0803i,3.3774$&$0.8199$        \\
$|a_{\{2,4,6\}}|$$\ll$$R$$\ll$$|a_0|$                &$2$&$4$&$0.7819$&$2$&$2$         \\
$R$$\gg$$|a_{\{0,2,4,6\}}|$     &$2$&$4^{(2)}$&$2^{(2)}$&$2$&$2$   \\
[0.05in]\hline\\
 {($f=3$)}  &$F_{\rm 3b}=6$ &$F_{\rm 3b}=7$ &$F_{\rm 3b}=8$ &$ F_{\rm 3b}=9$ \\
 Relevant $a_{\Ftb}$&$\{a_4,a_6\}$&$\{a_4,a_6\}$&$\{a_6\}$&$\{a_6\}$\\
[0.05in]\hline    
$R$$\ll$$|a_{\{0,2,4,6\}}|$ &$1.0062i,2.1662$&$1.0062i,2.1662$&$2.1162$&$1.0062i,4.4653$                                         \\
$|a_0|$$\ll$$R$$\ll$$|a_{\{2,4,6\}}|$    &$1.0062i,2.1662$&$1.0062i,2.1662$&$2.1162$&$1.0062i,4.4653$                  \\
$|a_2|$$\ll$$R$$\ll$$|a_{\{0,4,6\}}|$    &$1.0062i,2.1662$&$1.0062i,2.1662$&$2.1162$&$1.0062i,4.4653$                     \\
$|a_4|$$\ll$$R$$\ll$$|a_{\{0,2,6\}}|$    &$1.1329$&$0.6372$&$2.1162$&$1.0062i,4.4653$                     \\
$|a_6|$$\ll$$R$$\ll$$|a_{\{0,2,4\}}|$    &$0.5842i,3.5329$&$0.5491$&$4$&$2$                      \\ 
$|a_{\{0,2\}}|$$\ll$$R$$\ll$$|a_{\{4,6\}}|$    &$1.0062i,2.1662$&$1.0062i,2.1662$&$2.1162$&$1.0062i,4.4653$                \\
$|a_{\{0,4\}}|$$\ll$$R$$\ll$$|a_{\{2,6\}}|$	 &$1.1329$&$0.6372$&$2.1162$&$1.0062i,4.4653$  		\\
$|a_{\{0,6\}}|$$\ll$$R$$\ll$$|a_{\{2,4\}}|$    &$0.5842i,3.5329$&$0.5491$&$4$&$2$                  \\
$|a_{\{2,4\}}|$$\ll$$R$$\ll$$|a_{\{0,6\}}|$    &$1.1329$&$0.6372$&$2.1162$&$1.0062i,4.4653$                              \\
$|a_{\{2,6\}}|$$\ll$$R$$\ll$$|a_{\{0,4\}}|$     &$0.5842i,3.5329$&$0.5491$& $4$&$2$                          \\
$|a_{\{4,6\}}|$$\ll$$R$$\ll$$|a_{\{0,2\}}|$      &$2$&$2$&$4$&$2$                              \\
$|a_{\{0,2,4\}}|$$\ll$$R$$\ll$$|a_6|$        &$1.1329$&$.6372$&$2.1162$&$1.0062i,4.4653$                 \\
$|a_{\{0,2,6\}}|$$\ll$$R$$\ll$$|a_4|$        &$0.5842i,3.5329$&$0.5491$&$4$&$2$                   \\
$|a_{\{0,4,6\}}|$$\ll$$R$$\ll$$|a_2|$        &$2$&$2$&$4$&$2$                \\
$|a_{\{2,4,6\}}|$$\ll$$R$$\ll$$|a_0|$        &$2$&$2$&$4$&$2$                    \\
$R$$\gg$$|a_{\{0,2,4,6\}}|$      &$2$&$2$&$4$&$2$   
\end{tabular}}
\end{ruledtabular}
\end{table*}

Figure \ref{Potentials_f=1} shows the three-body potentials for $f=1$ as a result of solving Eq.~(\ref{eq:schrodinger}) for the allowed values of $\Fthb$.  
The scattering lengths chosen illustrate only one of a pair of possibilities for the relative strengths of the scattering lengths.  The 
hyperradius is written in terms of the typical range of interatomic interactions, $r_{\rm vdW}$.  The totally symmetric $\Fthb=1,3$ states, 
which are the only states not suppressed by centrifugal barriers, are associated with a pair of attractive potentials for $R\ll\{a_0,a_2\}$. 
This pair is associated with $s_0\approx 1.0062i$, and allow for the coexistence of two families of Efimov states (represented in Fig.~\ref{Potentials_f=1} 
by the horizontal solid and dash-dotted lines.)  As the hyperradius exceeds $a_0$, the $\Fthb=1$ attractive potential becomes an atom-dimer channel for collisions between the dimer $|\Ftb=0,\Mftb=0\>$ with energy $-1/ma_0^2$ and a free atom in the $|m_f=0\>$ state.  In the intermediate region $a_0 \ll R\ll a_2$, only the Efimov potential in the $\Fthb=3$ state remains and is associated with $s_0\approx 1.0062i$.  For $a_2\ll R$, the $\Fthb=3$ Efimov potential joins repulsive barriers from $\Fthb=1,2$ in an atom-dimer channel $|\Ftb=2,m_f=-1,0,1\>|m_f=-1,0,1\>$, which introduces the interesting possibility of studying atom-dimer spin mixtures \cite{Colussi2014}.  

As a final note, if a small (but finite) field is applied, the Efimov states in $\Fthb=1$ and 3 can interact creating an 
overlapping series of states.  Such controllability can produce ultra-long-lived states \cite{wang1991PRA} whose presence 
can affect the spin dynamics of the condensate depending on the short-range physics \cite{JILA_Theo}.  In the case of higher 
spin analyzed in the following section, many families of Efimov states with additional novel roots appear, along with the presence 
of overlapping series of Efimov states in the {\it absence} of an applied field.  

\section{\label{sec:lvel4} Spin-$2$ and -$3$ Systems}

In this section we go beyond the analysis from Ref.~\cite{Colussi2014} and detail the three-body potentials and Efimov states for $f=2$ and $3$ 
spinor condensates.  For spin-2 there are three rotationally invariant scattering lengths $\{a_0,a_2,a_4\}$ and $F_{\rm 3b}=0...6$. 
The three-body spin states [Eq.~(\ref{eq:threebodyspinfunctions})] for $F_{\rm 3b}=0$, 2, 3, 4, and 6 are symmetric while 
for $F_{\rm 3b}=1$ and 5 they are of mixed-symmetry.  
Shown in Table I are the relevant roots $s_\nu$ and relevant scattering lengths for each $\Fthb$ state. 
Similar to the $f=1$ case, for $f=2$ the roots for the regions in which $R$ is smaller than all relevant scattering lengths 
correspond to the roots for the usual identical bosons problem. All other regions display new roots.
For instance, within the region $|a_4|\ll R\ll |a_{\{0,2\}}|$ novel imaginary roots $s_0\approx0.3788i$ ($\Fthb=2$) and $s_0\approx0.5528i$ ($\Fthb=3$) can
be found. Analogously to the discussion of the spin configurations in Section III, the presence of these roots can be explained by recognizing that 
the three-body spin functions are not simple product states. 

\begin{figure}[htbp]
\includegraphics[width=3.375in]{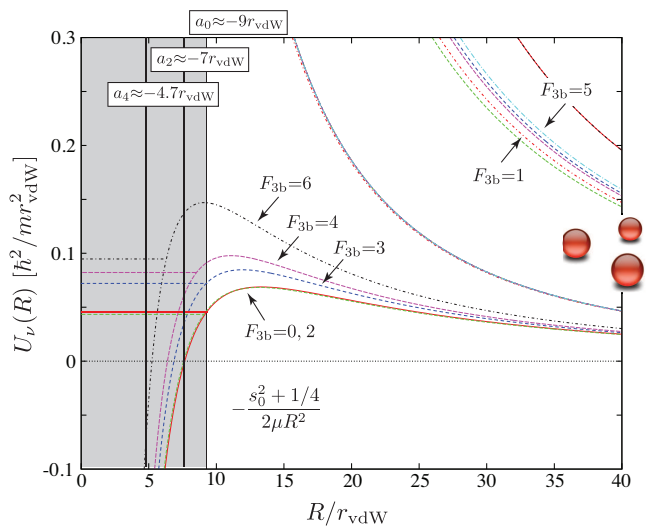}
\caption{$\Fthb=0$ (red solid lines), 1 (red dot-dashed lines), 2 (green dashed lines), 3 (blue dashed lines), 4 (purple dashed lines), 5 (cyan dot-dashed lines)
and 6 (black dot-dashed lines) hyperspherical adiabatic potentials for $f=2$ $^{85}$Rb atoms with $a_0\approx-8.97r_{\rm vdW}$, $a_2\approx-6.91r_{\rm vdW}$, and $a_4\approx-4.73r_{\rm vdW}$.  Note that the potentials for the mixed-symmetry states, $F_{\rm 3b}=1$ and 5, do not display an Efimov potential.
\label{Potentials85Rb}}
\end{figure}

Although typically the scattering lengths 
for most alkali species are small (comparable to $r_{\rm vdW}$), one notable exception is $^{85}$Rb with $f=2$. For that system 
$a_0\approx-8.97r_{\rm vdW}$, $a_2\approx-6.91r_{\rm vdW}$, and $a_4\approx-4.73r_{\rm vdW}$ \cite{klausen2001PRA,kempen2002PRL}.
Recent experimental and theoretical advances with alkali atoms have determined that the value of the scattering length at which the first Efimov 
state appears is approximately $-10r_{\rm vdW}$ 
\cite{IBK_Exps,LENS_Exps,Rice_Exps,Kayk_Exps,Jochim_Exps,Ohara_Exps,Ueda_Exps,JILA_Exps,Chin3BP,Ueda_Theo,JILA_Theo,Schimdt,Jensen}.  
Therefore, for $f=2$ $^{85}$Rb spinor condensates, it is not unreasonable to expect that the Efimov physics can strongly impact the spin dynamics 
and many-body physics.  
In Fig. \ref{Potentials85Rb}, the hyperspherical adiabatic three-body potentials for spin-2 $^{85}$Rb are shown. 
As one can see, the potentials for $\Fthb=0$, 2, 3, 4 and 6 states support an attractive Efimov potentials for distances smaller than and comparable
to the relevant scattering lengths, but forming a repulsive barrier as the hyperradius exceeds those distances.  
The potentials for the mixed-symmetry states, $\Fthb=1$ and 5, are completely repulsive and support no Efimov states 

Finally, for three identical spin-3 atoms there are four rotationally invariant scattering lengths $\{a_0,a_2,a_4,a_6\}$  and $F_{\rm 3b}=0,...,9$. For spin-3,
the $F_{\rm 3b}=0$ spin state is fully anti-symmetric, while $F_{\rm 3b}=2$ and 8 states are mixed-symmetry states and all others being fully 
symmetric states. Candidates for interesting Efimov physics (i.e., some of the relevant scattering lengths are large)
are $^{85}$Rb, $^{133}$Cs and $^7$Li.  
In Table I, the values of $s_\nu$ for all values of $\Fthb$ and all hyperradial ranges are presented for the first time. 
Where the spin-2 analysis produced only a few novel imaginary values of $s_\nu$, there are many more families Efimov states 
for spin-3. Of particular note is the appearance of overlapping Efimov series for $\Fthb=3$ when $R\ll |a_{\{0,2,4,6\}}|$ and 
$|a_{\{6\}}\ll R\ll |a_{\{0,2,4\}}|$. Even at zero applied field these overlapping states have the potential of being ultra-long lived 
\cite{wang1991PRA} depending on the short-range physics \cite{JILA_Theo}.  
Analysis of these overlapping states is beyond the scope of the present work.  The final section concludes with an analysis of the 
impact of Efimov physics on the three-body observables and many-body physics.

\section{\label{sec:level5}Two- and three-body mean-field parameters}

In this section we discuss the important two- and three-body mean-field parameters for spinor condensates and their connection
to Efimov physics. It is important to notice that in typical spinor condensate experiments the system is prepared at some small but finite value of the magnetic
field. In that case, the Zeeman interaction breaks the degeneracy of the atomic levels and atoms can be prepared in a single hyperfine substate, typically
$m_f=0$ \cite{SpinorReview}. After this preparation state, the magnetic field is quickly ramped off making now all atomic levels degenerate
and, consequently, making other hyperfine substates energetically available. 
As a result, the system evolves from a state where all atoms are in the $m_f=0$ to a more complicated mixture of spins.
Evidently, this spin dynamics is primarily controlled by the interatomic interactions since the initial state is not an eigenstate 
of the total hyperfine angular momentum.

Therefore, in order to connect with many-body treatments and identify the important interaction terms that control the spin dynamics it is 
necessary to rewrite the rotational invariant interactions in a suitable form. One such way was elaborated in Refs.~\cite{SpinorTheo1,SpinorTheo1x}
for two-body interactions. In this case, the two-body scattering length operator $\hat{A}$ [see Eq.~(\ref{eq:SpinoneAmatrix})] can be written as
\begin{align}
\hat{A}=\sum_{\Ftb}a_{\Ftb}{\mathrm P}_{\Ftb},\label{eq:AmatrixRewrite}
\end{align}
where ${\mathrm P}_{\Ftb}=\sum_{M_{F_{\rm 2b}}}|\Ftb\Mftb\>\<\Ftb\Mftb|$ are projection operators into a two-body 
total hyperfine state. In Ref.~\cite{SpinorTheo1} it was shown that the relation,
\begin{equation}
\left(\vec{f}_1\cdot\vec{f}_2\right)^n=\sum_{\Ftb}\left[\frac{\Ftb\left(\Ftb+1\right)}{2}-f\left(f+1\right)\right]^n{\mathrm P}_{\Ftb}\label{eq:constraints}
\end{equation}
allows us to rewrite $\hat{A}$ as
\begin{equation}
\hat{A}=\sum_{n=0}^{f} \alpha_{2\mathrm b}^{(n)}\left(\vec{f}_1\cdot\vec{f}_2\right)^{n},\label{eq:AMatrixRewrite2}
\end{equation}
where the $\alpha_{2\mathrm b}$'s are linear combinations of the scattering lengths $\{a_0,a_2,...,a_{2f}\}$.  

The above formulation allows us to rewrite the scattering length operator $\hat{A}$ in Eq.~(\ref{eq:AMatrixRewrite2}) 
for $f=1$ atoms in terms of two 
parameters, 
\begin{equation}
\hat{A}=\alpha_{2\mathrm b}^{(0)}+\alpha_{\rm 2b}^{(1)}\left(\vec{f}_1\cdot\vec{f}_2\right),\label{eq:spin1mf}
\end{equation}
where
\begin{eqnarray}
\alpha_{2\mathrm b}^{(0)}=\frac{a_0+2a_2}{3}\mbox{~~~and~~~}
\alpha_{2\mathrm b}^{(1)}=\frac{a_2-a_0}{3}, \label{A2bf=1}
\end{eqnarray}
representing a direct interaction terms and an spin-exchange term, respectively. 
It has been shown in Refs. \cite{SpinorTheo1,SpinorTheo1x} that these two parameters do characterize important phases of the gas.
For instance, depending on the sign of $\alpha_{2\mathrm b}^{(1)}$, the ground state of the spinor condensate is 
antiferromagnetic ($\alpha_{2\mathrm b}^{(1)}<0$) or ferromagnetic ($\alpha_{2\mathrm b}^{(1)}>0$) \cite{SpinorReview,
SpinorTheo1,SpinorTheo1x}. For $f=2$, the scattering length operator $\hat{A}$ in Eq.~(\ref{eq:AMatrixRewrite2}) 
can be rewritten in terms of three parameters
\begin{align}
&\alpha_{2\mathrm b}^{(0)}=-\frac{2a_0}{5}+\frac{8a_2}{7}+\frac{9a_4}{35}, \nonumber\\
&\alpha_{2\mathrm b}^{(1)}=-\frac{a_0}{30}-\frac{2a_2}{21}+\frac{9a_4}{70}, \nonumber\\ 
&\alpha_{2\mathrm b}^{(2)}=\frac{a_0}{30}-\frac{a_2}{21}+\frac{a_4}{70}.
\end{align} 
We note that in the literature the $\hat{A}$ operator for $f=2$ has been rewritten in terms of the projector 
${\mathrm P_0}$ allowing for a structure similar to $d$-wave BCS superfluids \cite{HoSpin2}. 
For $f=2$, there are three categories of ground state phases depending on the relative strengths of the scattering 
lengths: cyclic, polar, and ferromagnetic (see Ref.~\cite{HoSpin2}.)  
The cyclic phase arises from the additional ${\mathrm P_0}$ term which describes scattering of a pair of atoms into the $\Ftb=0$ 
singlet state introducing an additional order parameter describing the formation of singlet pairs in the system.  
For $f=3$, the four  parameters that characterize $\hat{A}$ [Eq.~(\ref{eq:AMatrixRewrite2})] are: $\alpha_{2\mathrm b}^{(0)}=9a_0/35-4a_2/7+486a_4/385+4a_6/77$, 
$\alpha_{2\mathrm b}^{(2)}=9a_0/70-17a_2/63+81a_4/770+25a_6/693$, $\alpha_{2\mathrm b}^{(4)}=-a_0/315+5a_2/378-6a_4/385+23a_6/4158$, 
and $\alpha_{2\mathrm b}^{(6)}=-a_0/630+a_2/378-a_4/770+a_6/4158$.  $\hat{A}$ was rewritten in Ref.~\cite{Spin3Mott} including 
${\mathrm P}_0$ and rewritten in Ref.~\cite{HoSpin3} including a nematic tensor for the dipolar spinor condensate $^{52}$Cr using 
the terminology of liquid crystals for the many ground state phases of the system at zero and nonzero fields which we do not go 
into further

There are, however, many ways to rewrite $\hat{A}$ in the literature that better suggest treatments similar, for instance, 
to magnetism, BCS theory, and liquid crystal theory. For the three-body mean-field contributions, however, we choose to 
simply extend the above analysis and in an analogous form to  Eq.~(\ref{eq:AMatrixRewrite2}), in light of the lack of suggestive 
three-body treatments from other theories to our knowledge. Therefore, we define the three-body scattering length 
operator as
\begin{equation}
\hat{A}_{{\rm 3b}}=\sum_{\Fthb}a_{\rm 3b}^{(F_{\rm 3b})}{\cal P}_{F_{\rm 3b}},
\end{equation}
where 
\begin{eqnarray}
{\cal P}_{F_{\rm 3b}}=\sum_{M_{F_{\rm 3b}}F_{\rm 2b}}
|F_{\rm 3b}M_{F_{\rm 3b}}(F_{\rm 2b})\rangle \langle F_{\rm 3b}M_{F_{\rm 3b}}(F_{\rm 2b})|.
\end{eqnarray}
In the equation above, $a_{\rm 3b}$ is the 
usual three-body scattering length (units of length$^4$) as defined in Refs. \cite{efimov1970SJNP,BraatenNieto1999,BraatenHammerMehen2002,BedaqueBulgacRupak2003,bulgac2002PRL}
for the allowed values of $F_{\rm 3b}$. Now, using the usual relations 
for angular momentum addition and the orthogonality of the projection operators, one can obtain the following relation,
\begin{equation}
\left(\sum_{i<j}{\vec{f}_i\cdot\vec{f}_j}\right)^n=\sum_{F_{\rm 3b}}\left[\frac{{F}_{\rm 3b}({F}_{\rm 3b}+1)}{2}-\frac{3f(f+1)}{2}\right]^n {\cal P}_{F_{\rm 3b}},
\end{equation}
which allows us to write the three-body scattering length operator $\hat{A}_{\rm 3b}$ as
\begin{equation}
\hat{A}_{\rm 3b}=\sum_{n=0}^{N_{3\mathrm{b}}-1}\alpha_{3\mathrm{b}}^{(n)}\left(\sum_{i<j}{\vec{f}_i\cdot\vec{f}_j}\right)^n,\label{A3bFull}
\end{equation}
where $N_{3\mathrm{b}}$ is the number of relevant three-body scattering lengths, i.e., the three-body scattering length for the values of
$F_{\rm 3b}$ whose spin functions are fully symmetric. (Anti-symmetric and mixed-symmetry states correspond to higher partial waves and are not 
considered in the present work.)  
Therefore, the three-body mean-field contributions to the spin dynamics is now determined in terms of the parameters 
$\alpha_{\rm 3b}$, which are linear combination of the physical three-body scattering lengths for all allowed values of $F_{\rm 3b}$.

For $f=1$ atoms, only the three-body scattering lengths for $\Fthb=1$ and $3$ contribute to the interaction (see Table I), allowing us to
write the three-body scattering length operator $\hat A_{\rm 3b}$ in Eq.~(\ref{A3bFull}) in terms of only two parameters,
\begin{eqnarray}
\hat{A}_{\rm 3b}=\alpha_{\rm 3b}^{\rm (0)}+\alpha_{\rm 3b}^{\rm (1)} \left(\sum_{i<j}{\vec{f}_i\cdot\vec{f}_j}\right),\label{SA3bS}
\end{eqnarray}
where
\begin{align}
\alpha_{\rm 3b}^{\rm (0)}=\frac{3a_{\rm 3b}^{(1)}+2a_{\rm 3b}^{(3)}}{5},\mbox{~~~and~~~}
\alpha_{\rm 3b}^{\rm (1)}=\frac{a_{\rm 3b}^{(3)}-a_{\rm 3b}^{(1)}}{5}, \label{A3b_f=1}
\end{align}
representing a direct interaction term and an spin-exchange term, respectively. This is in close analogy to the form
of the two-body scattering length operator for $f=1$ atoms [see Eq.~(\ref{eq:spin1mf})].
For $f=2$ atoms, only the states with $F_{\rm 3b}=0$, 2, 3, 4, and 6 contribute in the interaction, resulting in the following parameters
for Eq.~(\ref{A3bFull}),
 \begin{align}
&\alpha_{{\rm 3b}}^{(0)}=\frac{2 a_{3{\rm b}}^{(0)}}{35}-\frac{2 a_{3{\rm b}}^{(2)}}{7}+\frac{3 a_{3{\rm b}}^{(3)}}{5}+\frac{243
   a_{3{\rm b}}^{(4)}}{385}-\frac{a_{3{\rm b}}^{(6)}}{385},\nonumber\\
& \alpha_{{\rm 3b}}^{(1)}=  -\frac{a_{3{\rm b}}^{(0)}}{30}+\frac{23 a_{3{\rm b}}^{(2)}}{126}-\frac{29 a_{3{\rm b}}^{(3)}}{60}+\frac{513
   a_{3{\rm b}}^{(4)}}{1540}+\frac{a_{3{\rm b}}^{(6)}}{990},\nonumber\\
&  \alpha_{{\rm 3b}}^{(2)}= -\frac{29 a_{3{\rm b}}^{(0)}}{1260}+\frac{13 a_{3{\rm b}}^{(2)}}{126}-\frac{43 a_{3{\rm b}}^{(3)}}{360}+\frac{117
   a_{3{\rm b}}^{(4)}}{3080}+\frac{a_{3{\rm b}}^{(6)}}{770},\nonumber\\
&\alpha_{{\rm 3b}}^{(3)}=   -\frac{a_{3{\rm b}}^{(0)}}{945}+\frac{a_{3{\rm b}}^{(2)}}{1134}+\frac{a_{3{\rm b}}^{(3)}}{540}-\frac{3
   a_{3{\rm b}}^{(4)}}{1540}+\frac{17 a_{3{\rm b}}^{(6)}}{62370},\nonumber\\
&\alpha_{{\rm 3b}}^{(4)}=   \frac{a_{3{\rm b}}^{(0)}}{3780}-\frac{a_{3{\rm b}}^{(2)}}{1134}+\frac{a_{3{\rm b}}^{(3)}}{1080}-
\frac{a_{3{\rm b}}^{(4)}}{3080}+\frac{a_{3{\rm b}}^{(6)}}{62370}.
\end{align}
For $f=3$ atoms, the degree of complexity increases rapidly where now 
the states with $F_{\rm 3b}=1$, 3, 4, 5, 6, 7, and 9 contribute in the interaction (see Table I) and analog expressions for 
$\alpha_{\rm 3b}$ can be obtained.

In order to estimate the two- and three-body mean-field contributions and their relative importance,
we analyze the two- and three-body coupling constants
\begin{align}
g_{{\rm 2b}}^{(i)}=\frac{4\pi}{m}\alpha_{\rm 2b}^{(i)}\mbox{~~~and~~~}
g_{{\rm 3b}}^{(i)}=3^{1/2}\frac{12\pi}{m}\alpha_{{\rm 3b}}^{(i)}, 
\end{align}
respectively \cite{bulgac2002PRL}. This way, we can estimate the two- and three-body mean-field energies
simply as $n g_{\rm 2b}$ and $n^2g_{\rm 3b}$, respectively, with $n$ being the atomic density. 
This makes clear that relative importance of three-body effects will depend on the density (the denser
the gas the more important three-body contributions are.)  For instance, as pointed out in Ref.~\cite{Colussi2014},
the $f=1$ ferromagnetic and anti-ferromagnetic phases can be affected by three-body physics whenever
the three-body mean-field energy exceeds the two-body mean-field energy, i.e.,
\begin{eqnarray}
n^2 |g_{{\rm 3b}}^{(1)}|>n|g_{{\rm 2b}}^{(1)}|,
\end{eqnarray}
 and are of opposite sign. It is interesting noting that for most of the alkali species the two-body scattering 
lengths are typically small, implying that the two-body direct and spin-exchange mean-field energies are also
small opening up ways to explore the importance of three-body effects for the system. 
The fact that the two-body scattering lengths are small does not necessarily imply that the three-body scattering lengths, $a_{\rm 3b}$, 
are also small. A notable case is the $f=1$ $^{87}$Rb spinor condensate where not only are the scattering lengths small, 
but they are also approximately the same. For this case, $a_0\approx1.23r_{\rm vdW}$ and $a_2=1.21r_{\rm vdW}$ 
\cite{klausen2001PRA,kempen2002PRL} implying an extremely small spin-exchange energy term 
$\alpha_{\rm 2b}^{(1)}\approx-5.6\times10^{-3}r_{\rm vdW}$  [see Eq.~(\ref{A2bf=1})], and the three-body spin-exchange 
can be important. Determining the three-body contributions (the actual value of the three-body scattering lengths)
for such cases is extremely challenging since it would require a full numerical calculation including realistic two- and 
three-body interactions.

In the strongly interacting regime, not only do the three-body correlations become important, but they also become
universal \cite{Colussi2014}. The Efimov physics for the spin problem displays many features that can allow for
an independent control of both two- and three-body physics opening up ways to strongly modify the spin dynamics
in the condensate. From the analysis above, it is clear that the impact of Efimov physics to the spin dynamics is made 
through its role in three-body elastic processes. 
In general, Efimov states are manifested in the three-body scattering observables through log-periodic interference and resonant 
effects. In fact, for $f=1$ atoms (see Ref.~\cite{Colussi2014}), a simple WKB model  \cite{WKB} can be
used to determine the general form of the three-body scattering length. For ($f=1$), an interesting case emerges when both $a_0$ and 
$a_2$ assume large and negative values. In that case, if $|a_0|\gg|a_2|$ an analytical expression for the two relevant
three-body scattering lengths ($F_{\rm 3b}=1$ and $3$) can be determined and are given by
\begin{align}
a_{\rm 3b}^{(1)}&=\left[\alpha-\beta\tan\left(s_0\ln\frac{a_2}{a^-_{1}}\right)\right]\left(\frac{a_2}{a_0}\right)^{0.82}a_0^4+\gamma a_{0}^{4},\\
a_{\rm 3b}^{(3)}&=\left[\alpha-\beta\tan\left(s_0\ln\frac{a_2}{a^-_3}\right)\right]a_{2}^4,
\end{align}
where $s_0\approx1.0062$, and $\alpha$, $\beta$ and $\gamma$ are universal constants that can be determined
from numerical calculations for each value of $F_{\rm 3b}$. [The factor $(a_2/a_0)$ above originates from the regions
where the three-body potentials are repulsive ---see Table I.] For negative values of $a_0$ and $a_2$, the Efimov physics is manifested 
in elastic processes when a Efimov state becomes bound. In the above equations for $a_{F_{\rm 3b}}^-<0$  are the
values  $a_2<0$ where a $F_{\rm 3b}=1$ and $3$ Efimov resonance occurs. Near such values of $a_{F_{\rm 3b}}^-$, the
three-body scattering length $a_{\rm 3b}$ can be extremely large, assuming both positive and negative values, 
allowing for control of the spin dynamics via the spin-exchange term in Eq.~(\ref{A3b_f=1}).
In reality, since three-body losses are present and Efimov states have a finite lifetime, $a_{\rm 3b}$ does not strictly 
diverge and, in fact, is a complex quantity whose real and imaginary parts describe elastic and inelastic collisions---one can introduce loss effects in the expressions above (and below) by adding an imaginary phase term $i\eta$ in the argument of the tangent,
where $\eta$ is the so-called three-body inelasticity parameter \cite{braaten2006PR,wang2013AAMOP}.
 Nevertheless, for typical values of $\eta$ \cite{braaten2006PR,wang2013AAMOP} one can still expect substantial tunability
 of the three-body spin-exchange dynamics.

For $f=2$ atoms, evidently, the complexity in determining the three-body spin-dynamics increases. As we mentioned above,
an interesting case is presented for $f=2$ $^{85}$Rb atoms, where all the relevant two-body scattering lengths are large and negative
($a_0\approx-8.97r_{\rm vdW}$, $a_2\approx-6.91r_{\rm vdW}$, and $a_4\approx-4.73r_{\rm vdW}$ \cite{klausen2001PRA,kempen2002PRL})
and close to the value in which identical bosons display an Efimov resonance, $-10r_{\rm vdW}$ 
\cite{IBK_Exps,LENS_Exps,Rice_Exps,Kayk_Exps,Jochim_Exps,Ohara_Exps,Ueda_Exps,JILA_Exps,Chin3BP,Ueda_Theo,JILA_Theo,Schimdt,Jensen}.  
This implies that Efimov resonances might enhance three-body spin-exchange. Since the magnitude of the scattering lengths in this case are comparable,
it makes it difficult to obtain an accurate universal expression for the three--body scattering lengths. Nevertheless, it is instructive to analyze the results
obtained by assuming $|a_0|\gg|a_2|\gg|a_4|$. In this case, the relevant three-body scattering lengths are given by
 \begin{widetext}
\begin{align}
a_{\rm 3b}^{(0)}&=\left[\alpha-\beta\tan\left(s_0\ln\frac{a_2}{a^-_0}\right)\right]a_{2}^4\\
a_{\rm 3b}^{(2)}&=\left[\alpha-\beta\tan\left(s_0\ln\frac{a_4}{a^-_{2}}+s_0'\ln\frac{a_2}{a_4}\right)\right]\left(\frac{a_{2}}{a_{0}}\right)^{1.37}a_{0}^4+\gamma a_{0}^{4}\\
a_{\rm 3b}^{(3)}&=\left[\alpha-\beta\tan\left(s_0\ln\frac{a_4}{a^-_{3}}+s_0'\ln\frac{a_2}{a_4}\right)\right]a_{2}^4\\
a_{\rm 3b}^{(4)}&=\left[\alpha-\beta\tan\left(s_0\ln\frac{a_4}{a^-_{4}}\right)\right]\left(\frac{a_{4}}{a_{2}}\right)^{1.04}a_{2}^4+\gamma a_{2}^{4}\\
a_{\rm 3b}^{(6)}&=\left[\alpha-\beta\tan\left(s_0\ln\frac{a_4}{a^-_6}\right)\right]a_{4}^4
\end{align}
\end{widetext}
where $s_0\approx1.0062$, $s_0'\approx0.3788$, and $a_{F_{\rm 3b}}^-<0$ are the values of the two-body scattering lengths
where a $F_{\rm 3b}$ Efimov resonance occur. Note that now two distinct families of Efimov states (associated to $s_0$ and $s_0'$) 
can affect the spin dynamics. Although we expect the values for $a_{F_{\rm 3b}}^-$ to be close to -10$r_{\rm vdW}$ an analysis
similar to the one for identical bosons \cite{JILA_Theo} is necessary to precisely determine such values.

For higher spin the three-body contribution introduces additional complexity on the manifold of possible ground state phases of the spinor 
condensate.  The full study of the impact of three-body contributions on the many-body phases of spinor condensates remains however 
the subject of future investigation.  
 
\section{\label{sec:level6}Conclusion}

In summary, we have solved the three-body problem for strongly-correlated spinor condensates.  In this regime, the Efimov effect can 
play an important role for the spin dynamics of the condensate. For spinor systems multiple families of Efimov states can exist, leading to novel scaling 
laws and interesting spin dynamics. For small applied field, the families of Efimov states of different spin states can couple and may produce ultra long-lived 
states.  Curiously, in the case of spin-3 atoms there exist overlapping Efimov series even at zero applied field.  
The three-body hyperspherical adiabatic potentials for $^{85}$Rb are analyzed to illustrate possible strong effects due to
Efimov states. Finally, the existence of Efimov states can impact the three-body scattering observables and 
many-body phases of the system, and offer novel regimes for the spinor system.  

\section{Acknowledgements}

This work was supported by the U.S. National Science Foundation, grant numbers PHY-1125844, PHY-1307380 and PHY-1306905.

\end{document}